\documentclass[nolinenumbers]{aastex631}

\usepackage{txfonts}
\usepackage{comment} 
\usepackage{booktabs, threeparttable}
\usepackage{float} 

\shorttitle{AASTeX v6.3.1 Sample article}
\shortauthors{Bach-Møller et al.}

\graphicspath{{./}{figures/}}

\begin{document}

\title{Aggregation and charging of mineral cloud particles under high-energy irradiation}

\author[0000-0002-8799-0080]{Nanna Bach-Møller}
\affiliation{Centre for ExoLife Sciences, Niels Bohr Institute, Øster Voldgade 5, 1350 Copenhagen, Denmark}
\affiliation{Space Research Institute, Austrian Academy of Science, Schmiedelstraße 6, 8042 Graz, Austria}
\affiliation{Institute for Theoretical Physics and Computational Physics, Graz University of Technology, Petersgasse 16, 8010 Graz, Austria}

\author[0000-0002-8275-1371]{Christiane Helling}
\affiliation{Space Research Institute, Austrian Academy of Science, Schmiedelstraße 6, 8042 Graz, Austria}
\affiliation{Institute for Theoretical Physics and Computational Physics, Graz University of Technology, Petersgasse 16, 8010 Graz, Austria}

\author[0000-0001-7303-914X]{Uffe G. Jørgensen}
\affiliation{Centre for ExoLife Sciences, Niels Bohr Institute, Øster Voldgade 5, 1350 Copenhagen, Denmark}

\author[0000-0001-8452-698X]{Martin B. Enghoff}
\affiliation{DTU Space, Danish Technical University, Elektrovej 327, 2800 Kgs. Lyngby, Denmark}

\begin{abstract}

It is known from Earth that ionizing high-energy radiation can lead to ion-induced nucleation of cloud condensation nuclei in the atmosphere. Since the amount of high-energy radiation can vary greatly based on the radiative environment of a host star, understanding the effect of high-energy radiation on cloud particles is critical to understand exoplanet atmospheres. 
This study aims to explore how high-energy radiation affects the aggregation and charging of mineral cloud particles. 
We present experiments conducted in an atmosphere chamber on mineral SiO$_2$ particles with diameters of 50 nm. The particles were exposed to gamma radiation in either low-humidity (RH $\approx$ 20\%) or high-humidity (RH $>$ 50\%) environments. The aggregation and charging state of the particles were studied with a Scanning Mobility Particle Sizer.
We find that the single SiO$_2$ particles (N1) cluster to form larger aggregates (N2 - N4), and that this aggregation is inhibited by gamma radiation. We find that gamma radiation shifts the charging of the particles to become more negative, by increasing the charging state of negatively charged particles. Through an independent T-test we find that this increase is statistically significant within a 5\% significance level for all aggregates in the high-humidity environment, and for all except the N1 particles in the low-humidity environment. For the positively charged particles the changes in charging state are not within the 5\% significance level.
We suggest that the overall effect of gamma radiation could favor the formation of a high number of small particles over a lower number of larger particles.

\end{abstract}

\keywords{Exoplanet atmospheres (487) --- Planetary atmospheres (1244) --- Atmospheric clouds (2180) --- High-energy cosmic radiation (731) --- Gamma-rays (637)}

%

\section{Introduction} 

One of the greatest advancements in recent years of exoplanet research is our improved understanding of exoplanet atmospheres. A major step in this regard has been the recognition of the critical role cloud formation plays in the energy balance and chemistry of atmospheres. Exoplanets, as well as many Solar System objects, display a wide variety of cloud particles that differ greatly from the cloud formation observed on Earth. In addition to the various water, ammonia, and condensed hydrocarbon clouds, that are among the many cloud types observed in the atmospheres of Solar System objects (e.g. \cite{baines_storm_2009,brooke1998models,brown2002direct,gao_aerosols_2021,ohno2021haze,rages1992voyager,romani1988methane,sagan1992titan,sromovsky2011methane,wong2017photochemistry}), studies predict a wide range of mineral clouds in warmer exoplanet atmospheres (e.g. \cite{gao_aerosols_2021,helling_clouds_2020,helling_cloud_2022,marley1999reflected,seager2000theoretical}), which is believed to greatly influence our observations of these planets \citep{gao_aerosols_2021,helling_cloud_2022}. \\

One of the less understood aspects of cloud formation, is the effects of the radiation field from the host star and the influx of high-energy particles into the atmosphere. Exoplanets are exposed to two major sources of high-energy particles: stellar energetic particles, and galactic cosmic rays (reviewed in \cite{rodgers-lee_stellar_2021}). 
\textit{Stellar energetic particles} (SEPs, also known as stellar cosmic rays) are high-energy particles irradiated by the host star either during stellar flares, or as particles accelerated by the shock front of coronal mass ejections (\cite{wild1963solar,reames2013two, rodgers-lee_galactic_2020}).
\textit{Galactic cosmic rays} (GCRs) are high-energy particles originating from outside the system, that travel trough interplanetary space before they reach the planet.

Since younger and more active stars generally have strong magnetic fields and thereby more stellar flares (as reviewed by \cite{getman_x-ray_2021}), the radiative environment around e.g. M-dwarfs will be very different from that of the Solar System both in relation to SEPs and GCRs. Studies have found that the high magnetic activity of M-dwarfs causes them to release higher fluxes of SEPs \citep{rodgers-lee_stellar_2021,fraschetti_stellar_2019} due to the strong stellar winds. At the same time, the propagation of GCRs through the stellar wind has long been studied \citep{parker_passage_1965,jokipii_propagation_1971,rodgers-lee_galactic_2020}, and it has been found that host stars with strong magnetic fields can shield their planets from GCRs (as reviewed in e.g. \cite{nandy_solar_2021}). An exoplanet around an M-dwarf host will therefore have a different influx of high-energy particles, and in order to understand the atmospheres of these planets, we need to understand the effect of high-energy radiation. \\

High-energy particles have been shown to have a significant effect on the atmosphere as a whole. Based on models, it has been predicted that SEPs can observably alter the chemical composition of exoplanet atmospheres \citep{barth_moves_2021,chadney_effect_2017,venot_influence_2016}, and destroy potential ozone layers in Earth-like atmospheres \citep{segura_effect_2010}.
The effect of high-energy particles on cloud formation has been studied extensively for Earth's atmosphere, and it has been found that there is a significant correlation between the influx of high-energy particles and the degree of cloud formation \citep{dickinson_solar_1975,svensmark_variation_1997,marsh_low_2000,svensmark_atmospheric_2021}. As high-energy particles reach the atmosphere, they form so called \textit{cosmic ray showers}, where the particles interact with the gas under the release of gamma rays resulting in an ionization of the gas. The increased presence of ions has been found to stabilize and promote the growth of molecular clusters in Earth's atmosphere, leading to ion-induced nucleation of cloud particles \citep{lee_new_2019,wagner_role_2017}.
A series of studies \citep{svensmark_response_2013,enghoff_measurement_2017,svensmark_ion-cage_2020} have looked at ion-induced nucleation by imitating high-energy particles using gamma radiation. \cite{svensmark_response_2013} found that gamma irradiation increases the nucleation of H$_2$SO$_4$ clusters from the gas-phase, and induces the nucleated clusters to grow to 50 nm sizes. Using a similar experimental setup, \cite{enghoff_measurement_2017} found that gamma radiation changes the charging state of the particle population and that particles react differently to gamma radiation depending on their size and polarization. \cite{enghoff_measurement_2017} suggest that different ions participate in the charging of positively and negatively charged particles, and that the differences in the size and mobility of these ions will affect the charging of the particles by the gamma irradiation. 

The effect of high-energy radiation is known to also apply to particles very different from the ones participating in Earth-like cloud formation (e.g. metallic nanoparticles \cite{abedini_review_2013}), indicating a relevance to fields other than cloud formation. One such example could be the study of interplanetary dust. Interplanetary dust has long played an important role in solar system space mission \citep{grun_galileo_1992,jorgensen_distribution_2021}, and an improved understanding of the effect of the radiation environment on particles could be highly beneficial for future solar system missions, such as Comet Intercepter \citep{snodgrass_european_2019}, and for our understanding of dust in planetary systems as a whole. \\

So far, the effect of high-energy particles on cloud formation has mostly been studied for the initial molecular nucleation from the gas phase of volatile species characteristic to the atmosphere of Earth. The aim of this study is to expand on previous studies, by answering the question of how high-energy radiation affects the aggregation and charging of already formed mineral cloud particles. The mineral cloud particles investigated in this study, SiO$_2$, are relevant both as mineral cloud species on gaseous exoplanets, AGB Stars and Brown Dwarfs, and as cloud condensation nuclei in Earth-like atmospheres \cite{lee_dynamic_2016,helling_detectability_2006}.

The experimental setup used in this study is similar to the setup used previously by \cite{enghoff_measurement_2017} and \cite{motzkus_size_2013}, and is described in Sec. \ref{sec:methods}. In Sec \ref{sec:results} we present the results of: the general size distribution of the particles (Sec \ref{sec:size}), the effect of gamma radiation on the aggregation of the particles (Sec. \ref{sec:aggregation}), the effect of gamma radiation on the charging state of the particles (Sec. \ref{sec:charge}), and the effect of humidity on both charging and aggregation (Sec. \ref{sec:highhum}). In Sec. \ref{sec:discussion} the results are discussed in relation to their implication on the population of exoplanet atmospheres.

\section{Methods} \label{sec:methods}	

\begin{figure*}
    \centering
    \resizebox{0.8\textwidth}{!}
    {\includegraphics[trim={7.5cm 0.7cm 5.5cm 2.5cm}, clip]{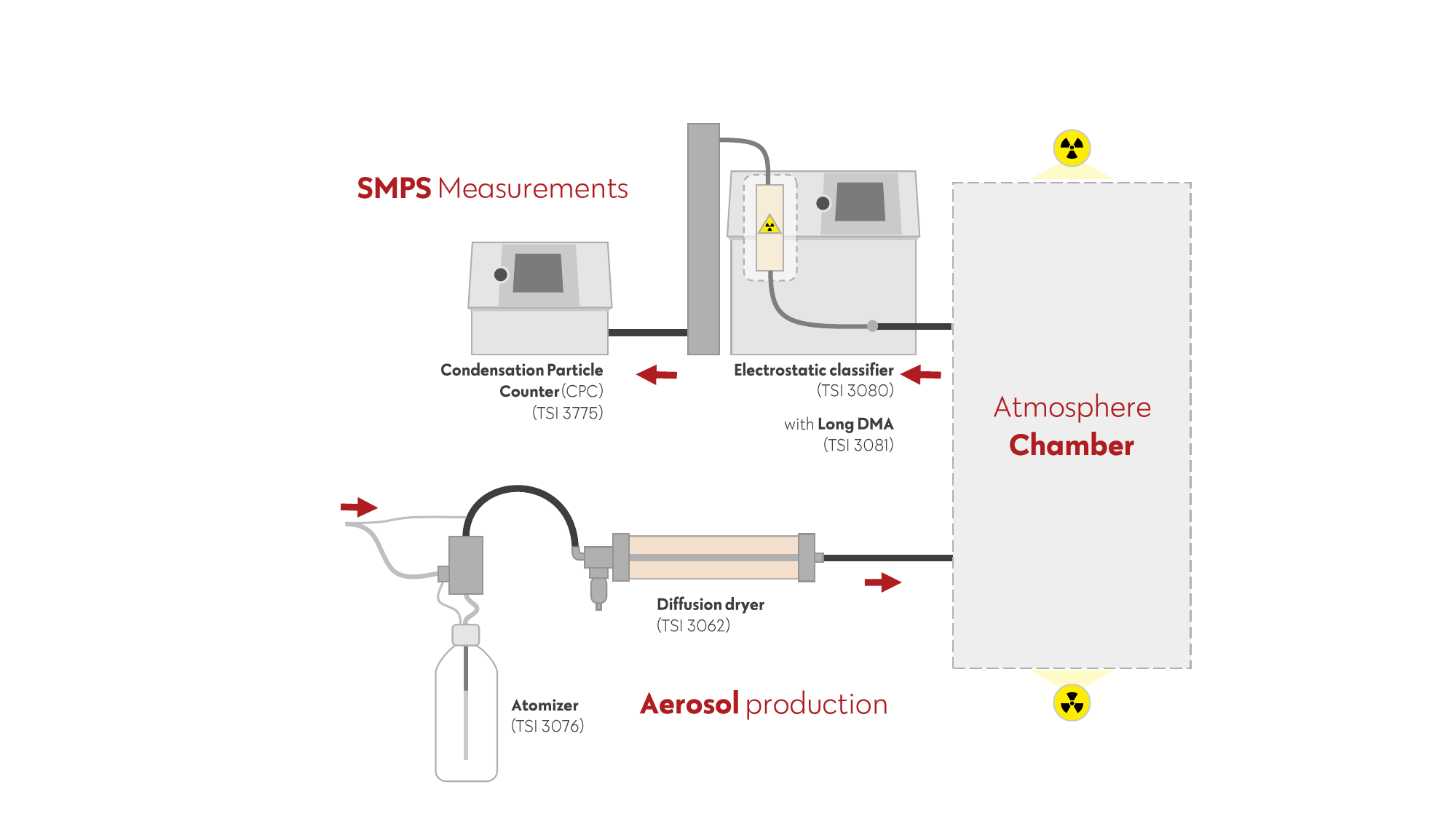}}
    \caption{Experimental setup. The process begins with aerosol production (lower left), continues to the atmosphere chamber (right), and ends with measurements at the SMPS system (upper left). Arrows mark the airflow through the system. }
    \label{fig:chamber}
\end{figure*}

The main purpose of the experimental setup is to address the question of how mineral cloud particles are affected by high-energy radiation. This is addressed by releasing mineral particles from a liquid solution and introduce them into an atmosphere chamber where they are irradiated with gamma radiation. The effect of the gamma radiation on the particle size and charging state is measured based on their mobility in a known electric field.
A single experiment in this study consists of 50 individual size distribution measurements, each with a scan time of 240 s, taken consecutively, corresponding to a total of $\sim$3.5 hours of measurements for each experiment.

\subsection{Setup}
The experimental setup can be separated into three parts: a) The production of aerosols (bottom left in Fig. \ref{fig:chamber}). b) The chamber (right in Fig. \ref{fig:chamber}). c) Measurements using a Scanning Mobility Particle Sizer, SMPS (top left in Fig. \ref{fig:chamber}),

\begin{table*}
\centering                          
\begin{tabular}{c c}        %
    \multicolumn{2}{c}{Universal settings for all experiment} \\
    \hline\hline                 
    Scan up time & 180 s \\
    Scan down time & 60 s \\
    Sheath flow & 3 L/min \\
    CPC flow & 0.3 L/min \\ 
    Multiple charge correction & ON \\
    Particle number density in atomizer & $\sim 7 \cdot 10^{9}$ mL$^{-1}$ \\
    Particle number density in chamber & $\sim 7 \cdot 10^{3}$ cm$^{-3}$ \\
    Total N$_2$ flow to chamber & 16 L/min \\
    \hline
\end{tabular}

\begin{threeparttable}
\begin{tabular}{c c c c c c c}        %
    &&&& \\
    \multicolumn{7}{c}{Specific settings for each experiment} \\
    \hline\hline                 
    Nr.\tnote{a} & Charge\tnote{b} & Gamma\tnote{c} & Bipolar charger\tnote{d} & RH & Duration\tnote{e} & Repetitions\tnote{f} \\    

\hline
    1 & Pos & No & Yes & $\sim 20\%$ & 3.5 & 6 \\
    2 & Pos & No & No & $\sim 20\%$ & 3.5 & 6 \\
    3 & Pos & Yes & Yes & $\sim 20\%$ & 3.5 & 6 \\
    4 & Pos & Yes & No & $\sim 20\%$ & 3.5 & 6 \\
    5 & Neg & No & Yes  & $\sim 20\%$ & 3.5 & 6 \\
    6 & Neg & No & No & $\sim 20\%$ & 3.5 & 6  \\
    7 & Neg & Yes & Yes & $\sim 20\%$ & 3.5 & 6 \\
    8 & Neg & Yes & No & $\sim 20\%$ & 3.5 & 6 \\
    9 & Neg & No & Yes & $\sim 20\%$ & 20 & 1 \\
    10 & Neg & Yes/No & Yes & $\sim 20\%$ & 30 & 1 \\
\hline
    &&&& \\
\hline               %
    11 & Pos & No & Yes & $\sim 65\%$ & 3.5 & 3 \\
    12 & Pos & No & No & $\sim 65\%$ & 3.5 & 3 \\
    13 & Pos & Yes & Yes & $\sim 65\%$ & 3.5 & 3 \\
    14 & Pos & Yes & No & $\sim 65\%$ & 3.5 & 3 \\
    15 & Neg & No & Yes  & $\sim 65\%$ & 3.5 & 3 \\
    16 & Neg & No & No & $\sim 65\%$ & 3.5 & 3  \\
    17 & Neg & Yes & Yes & $\sim 65\%$ & 3.5 & 3 \\
    18 & Neg & Yes & No & $\sim 65\%$ & 3.5 & 3 \\
    19 & Pos & Yes/No & Yes & $\sim 50\%$ & 22 & 1 \\
\hline                                   
\end{tabular}
\begin{tablenotes}
    {\footnotesize
    \item[a]Identifier for type of experiment.
    \item[b]Electric charge of measured particles.
    \item[c]Irradiation with gamma radiation.
    \item[d]Specifies whether measurements were done with the bipolar charger (yes) or the dummy (no)
    \item[e]Total duration of each experiment in hours.
    \item[f]Number of repetitions of experiments with these settings.
    }
\end{tablenotes}
\end{threeparttable}
\caption{Settings for experiments. Top part shows universal settings for all experiments. Bottom part shows experiment-specific settings for low-humidity (Nr. 1-10) and high-humidity (Nr. 11-19) environments respectively.} 
\label{tab:setup} 
\end{table*}

\subsubsection{Aerosol production}
The experiment is started by separating and releasing SiO$_2$ particles from a liquid solution, to form  non-interacting aerosols in an airflow with as low humidity as possible. 
The sample used is MSP NanoSilica\texttrademark; an aqueous suspension of amorphous SiO$_2$ particles with a highly uniform size distribution around diameters of 50 nm.

The sample is lowered into a sonic bath to ensure separation of the particles in the suspension, before it is diluted in Milli-Q water (water that has been purified and deionized using a Millipore Milli-Q lab water system) to a number density of $\sim 7 \cdot 10^{9}$ particles per mL.
The particles are released from the aqueous suspension by exposing the sample to a jet of compressed air using an atomizer (TSI 3076). This converts the SiO$_2$ water solution into an aerosol spray where the individual particles are moving, well separated, with the air flow. 

The atomizer is operated in re-circulation mode using a flow of clean dry nitrogen at a rate of 3 L/min. To decrease the humidity of the air flow, the flow from the atomizer is diluted with an extra flow of 1 L/min nitrogen and led through a diffusion dryer (TSI 3062), lowering the relative humidity from $\sim 100\%$ to $\sim 60\%$.

\subsubsection{Atmosphere chamber}
The aggregation and gamma irradiation of the particles takes place in the atmosphere chamber, where the aerosol flow is introduced with an additional airflow of 12 L/min clean nitrogen from a Parker Midigas 6 nitrogen generator with a O$_2$ content of max. 10 ppm. The lowest relative humidity in the chamber from the combined air flows from the nitrogen generator and atomizer is $\sim 20 \%$. The humidity can be further increased by passing the nitrogen air through a humidifier. The final gas composition of the chamber is therefore N$_2$ and H$_2$O. The atmosphere chamber has the dimensions $2m\times 2m\times 2m$, and is primarily made of stainless steel (with the exception of one side made of teflon to allow for experiments with UV). All experiments in this study are conducted at approximate room temperature (around 21-23 $^{\circ}$C) and slightly above atmospheric pressure (differential pressure of $\sim$0.2 mbar to the surroundings) to prevent the inflow of contaminating air from outside the chamber. The chamber has previously been described by \cite{svensmark_response_2013}. Two gamma sources (27 MBq Cs-137) are located on opposite sides of the chamber. The level of irradiation can be controlled by varying the shielding of the radiation sources, resulting in ion production rates ranging from 16-200 cm$^{-3}$s$^{-1}$. \\
Introducing the particles into the chamber at a steady rate, it takes approximately 10 hours for the particle number density to reach a steady state. All measurements are therefore taken 15 hours or more after the first introduction of the particles. Similarly, if there was a change in the irradiation of the chamber, at 30 minutes or more were allowed to pass before further measurements were taken, in order for the ionization to be allowed to take place. Over time particles will leave the gas phase through collision with the chamber walls which happens with a rate of $<$0.05 per hour, see appendix \ref{app:loss}.

\subsubsection{SMPS Measurements} \label{sec:SMPS}
In order to determine the potential changes in particle size and charge distribution the electric mobility distribution of  the particles is measured. The measurements are done using a Scanning Mobility Particle Sizer (SMPS) system consisting of an Electrostatic Classifier and a Condensation Particle Counter (CPC), as seen in the upper left of Fig. \ref{fig:chamber}. 

The principle behind the SMPS system is the correlation between the size and charge of the particles and their electrical mobility. The electrical mobility ($Z_p$) reflects the ability of a particle to move through an electric field as described by:
\begin{equation}
    Z_p = \frac{neC}{3\pi \mu D_p}
    \label{eq:mob}
\end{equation}
Where $n$ is the number of elementary charges on the particle, $e$ is the elementary charge, $C$ the Cunningham slip correction (that is depending on the gas mean free path which is calculated continuously and $D_p$), $\mu$ the gas viscosity, and $D_p$ the particle diameter \citep{tsi2009classifier}. For small particles or aggregated clusters the particle cannot be assumed to be spherical and $D_p$ will instead be defined as the \textit{mobility diameter}. The relationship between the electrical mobility and the parameters of the Classifier has been determined to \citep{knutson_aerosol_1975}:
\begin{equation}
    Z^*_p = \frac{q_{sh}}{2\pi \bar{V}L}ln\left(\frac{r_2}{r_1}\right)
    \label{eq:mob2}
\end{equation}

Where Z$^*_p$ is a set mobility, q$_{sh}$ is the sheath flow rate, $\bar{V}$ is the average voltage used for the DMA, L is the distance between the exit slit and the aerosol inlet in the DMA, and r$_1$ and r$_2$ are the inner and out radius of the annular space in the DMA.

Combining Eq. \ref{eq:mob} and Eq. \ref{eq:mob2}, the particle diameter can be related to the number of charges on the particles, the collector rod voltage, the flow rate of the classifier, and the geometry of the DMA \citep{tsi2009classifier}:

\begin{equation}
    \frac{D_p}{C} = \frac{2ne\bar{V}L}{3 \mu q_{sh} ln\left(\frac{r_2}{r_1}\right)}
    \label{eq:dia_charge}
\end{equation}

As can be seen from Eq. \ref{eq:dia_charge} the mobility diameter of a particle can be found based on the charge of the particle and the instrumental setup of the classifier.
The electrostatic classifier obtains a known charging distribution using a Kr-85 Bipolar Charger (also called a neutralizer). The bipolar charger ionizes the air and thereby expose the particles to high concentrations of bipolar ions. Through collisions with these ions the particles themselves are brought into a known steady-state bipolar charge distribution, where a fixed percentage of the particles will carry none, one, or multiple charges.
The charged particles are passed into a Long Differential Mobility Analyzer (DMA) (TSI 3081), where an electric field allows only particles in a specific electric mobility range to pass through. As the electric field is varied the particles are being sorted based on their mobility and and the number density of particles within each mobility range is measured by a Condensation Particle Counter (CPC) (TSI 3775). 
Accounting for the known charge distribution implemented by the bipolar charger \citep{wiedensohler_approximation_1988} this gives us a mobility diameter distribution of the particle population. The Electrostatic classifier is operated with a sheath flow of 3 Lmin$^{-1}$, the CPC was operated on low-flow mode at 0.3 Lmin$^{-1}$, and the long DMA was set to a size range of 13-500 $nm$. Scan times were chosen for the SMPS based on the flow rates and desired particle size range in addition to the CPC response time. The scan up time, determining the time spent by the classifier to increase the voltage over the DMA, was set to 180s, whereas the scan down time, determining the time taken by the classifier to return to the initial voltage, was set to 60s. This scan time of a total of 240s is in accordance with the settings previously used by \citep{motzkus_size_2013} to measure SiO$_2$ particles using a similar setup. The final size distributions are found by averaging over 50 consecutive scans as explained in Sec. \ref{sec:evaluation}. \\

In order to study the charging of the SiO$_2$ particles the charging state is calculated. The charging state is a measure of the charge of the particle population in relation to the steady-state charge distribution implemented by the Kr-85 bipolar charger \citep{laakso_detecting_2007}. This is done by replacing the neutralizing Kr-85 charger with a non-neutralizing "dummy", that is identical to the charger but with no Kr-85 source (as explained in \cite{enghoff_measurement_2017}). While using the "dummy" the particles will maintain their original charges from the chamber, and the measurements will reflect these charges in addition to the particle sizes. The charging state is calculated by dividing measurements performed with the dummy with identical measurements performed with the Kr-85 bipolar charger.

Since the electric field of the DMA will sort for either positively or negatively charged particles, these will be measured separately in the SMPS system. The polarity of the particles being measured is decided by changing the power supply in the classifier, such that a positive power supply will result in the negatively charged particles being measured, and a negative power supply will result in the positively charged particles being measured \citep{enghoff_measurement_2017}.

\subsection{Evaluation of measurements}\label{sec:evaluation}

All results in this study are based on size distribution measurements done with the SMPS. Based on these measurements we calculate: average size distributions of different "experiments", number density of different aggregates, and charging states, all with associated uncertainties.

\textit{Size distribution of experiment:}
A single experiment in this study consists of 50 individual size distribution measurements (scans) taken consecutively by the SMPS. The number density for each measurement is found from the normalized concentration measured by the SMPS (nN/dlogD$_p$) by multiplying it with the difference in the log of the size bin width. The final size distribution of each experiment is found as the average of the 50 consecutive measurements. The standard deviation of each experiment was found from the standard error of the mean for each size bin.

\textit{Number density of aggregates:} As shown in Fig. \ref{fig:size_both} and explained in Sec. \ref{sec:size}, the size distribution shows distinct peaks, indicating aggregates of different sizes. The number of aggregates of each size is found as the sum of measured particles within the full width half maximum (FWHM) of the peak. The uncertainty is found from through linear error propagation from the the standard deviation of each size bin in the size distribution of the experiment.

\textit{Charging state:}
In order to calculate the charging state, two experiments are performed after each other, one with the K-85 bipolar charger and one with the dummy (e.g. experiment 1 and 2 in Tab. \ref{tab:setup}). The number density of each aggregate size for each of the experiments is calculated, and the charging state is found as the number of particles from the measurement with the dummy divided by the measurement with the bipolar charger. A charging state of 1.0 would indicate that the aggregates in the chamber follow the same steady-state charge distribution as implemented by the bipolar charger. Charging states different from 1.0 indicate that the particles deviate from the steady-state distribution, thereby indicating that the aggregates are actively being charged. Values above 1.0 indicate an overcharge of the particles (a higher number of charged particles than in the steady-state), and values below 1.0 indicate an undercharge of the particles (a lower number of charged particles than in the steady-state).
The uncertainty of each charging state calculation is found using the standard formula for error propagation from the uncertainties of the particle number densities. For each of the four settings (positively or negatively charged particles, with or without gamma-irradiation), the charging states for each of the aggregates are averaged among the repeated experiments. The uncertainties for the averages are found in two ways: 1) Error propagation of the errors of each charging state. 2) The standard deviation among the repeated experiments.

The error bars shown in the figures indicate one standard deviation.

\section{Results} \label{sec:results}
In this section we show the results for the charging state and aggregation of positively and negatively charged particles respectively. The results have been obtained through at number of repeated experiments, all listed in Tab. \ref{tab:setup}. The results presented in Sec. \ref{sec:size} - \ref{sec:charge} are all done in low humidity environments (RH $\approx$ 20\%), whereas the experiments in Sec. \ref{sec:highhum} are done at higher humidity (RH $>$ 50\%).

\subsection{Size distribution} \label{sec:size}

\begin{figure}
    \centering
    \includegraphics[width=0.7\linewidth]{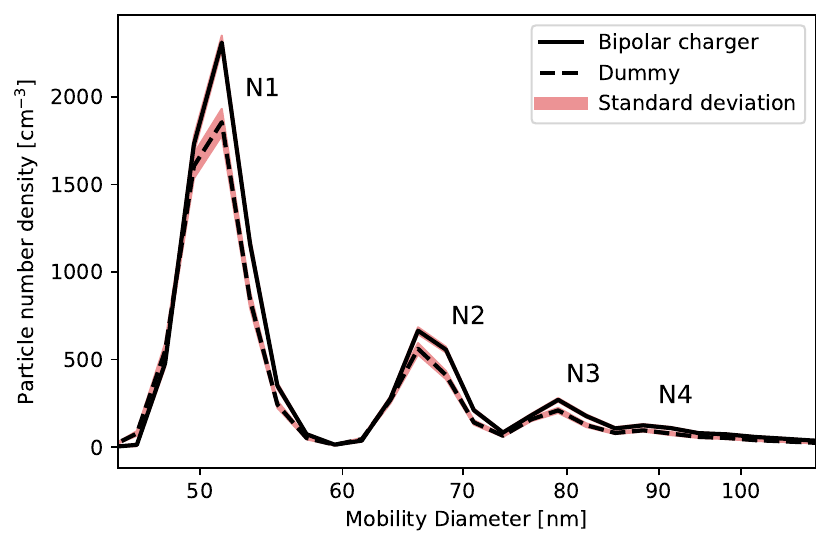}
    \caption{Size distribution of negatively charged particles measured with and without bipolar charger. Red shading indicates standard deviation among the 50 measurements constituting each experiment. Shows one repetition of experiment nr. 5 in Tab. \ref{tab:setup}.}
    \label{fig:size_both}
\end{figure}

\begin{figure}
    \centering
    \includegraphics[width=0.7\linewidth]{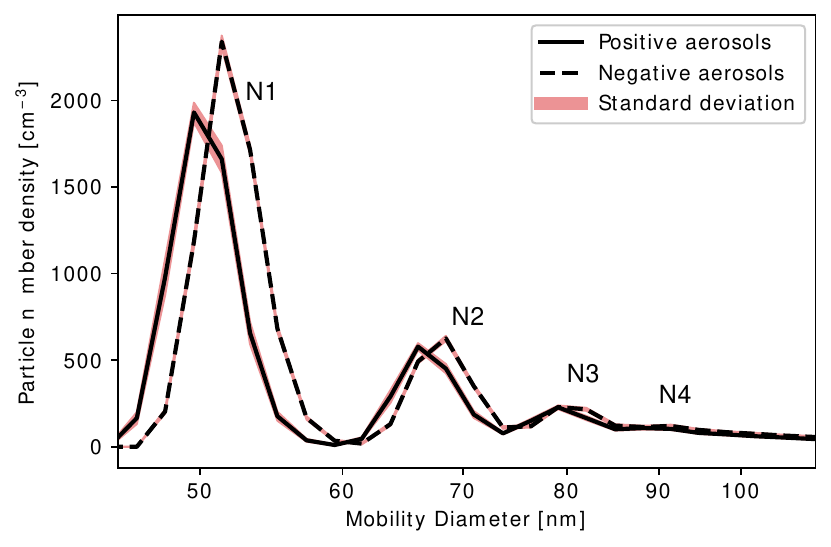}
    \caption{Comparison of size distribution of positively and negatively particles respectively. Both measurements are done with the bipolar charger. Red shading indicates standard deviation among the 50 measurements constituting each experiment. Shows one repetition of experiment nr. 5 and 6 in Tab. \ref{tab:setup}.}
    \label{fig:size_pn}
\end{figure}

An example of the size distributions measured by the SMPS can be seen in Fig. \ref{fig:size_both}. The two size distributions show experiments with and without the bipolar charger for the negatively charged particles. Measurements with the bipolar charger illustrate the actual size distribution of the particles, while measurements without the charger (dummy) also reflect the inherent charging of the particles. 

The size distributions show 3-4 distinct peaks corresponding to the original particles (with diameters of d $\approx$ 50 nm) and aggregates of two, three and four particles (i.e. N1, N2, N3, and N4). The mobility diameters of the aggregates, as shown on the primary axis, do not correspond exactly to physical diameter, but rather describe the mobility of the aggregates through the electric field of the DMA, which depends on the physical diameter as described in Sec. \ref{sec:SMPS}. The plots in Fig. \ref{fig:size_both} and \ref{fig:size_pn} show only the relevant mobility diameter range for the SiO$_2$ aggregates and thereby exclude features at lower mobility diameters caused by e.g. the Milli-Q water. By comparing the distribution with and without the bipolar charger for these peaks we see that more particles are being measured with the charger, indicating that there are fewer negatively charged particles in the population than we would expect from the steady-state bipolar charge distribution. This will be further discussed in Sec. \ref{sec:charge}.

The particles get their positive or negative charge from collisions with ions in the gas phase. Since the positive and negative ions can have different mobility \citep{horrak_bursts_1998}, the ions with the highest mobility will participate in more collisions, which can affect both the number of charged particles and their mobility through the DMA in the SMPS system. Fig. \ref{fig:size_pn} illustrates the size distribution of the positively and negatively charged particles respectively. As can be seen the distributions differ slightly, both in regard to the particle count and the position of the peaks. The measurements show a higher number of negatively charged particles at around 50 nm. We also see that the distribution for the negatively charged particles is shifted slightly towards higher diameters, indicating that their mobility in the DMA is lower than for the positively charged particles. These differences are not a major concern as charging state calculations are relative. \\

\subsection{Aggregation over time}\label{sec:aggregation}

\begin{figure}
    \centering
    \includegraphics[width=0.7\linewidth]{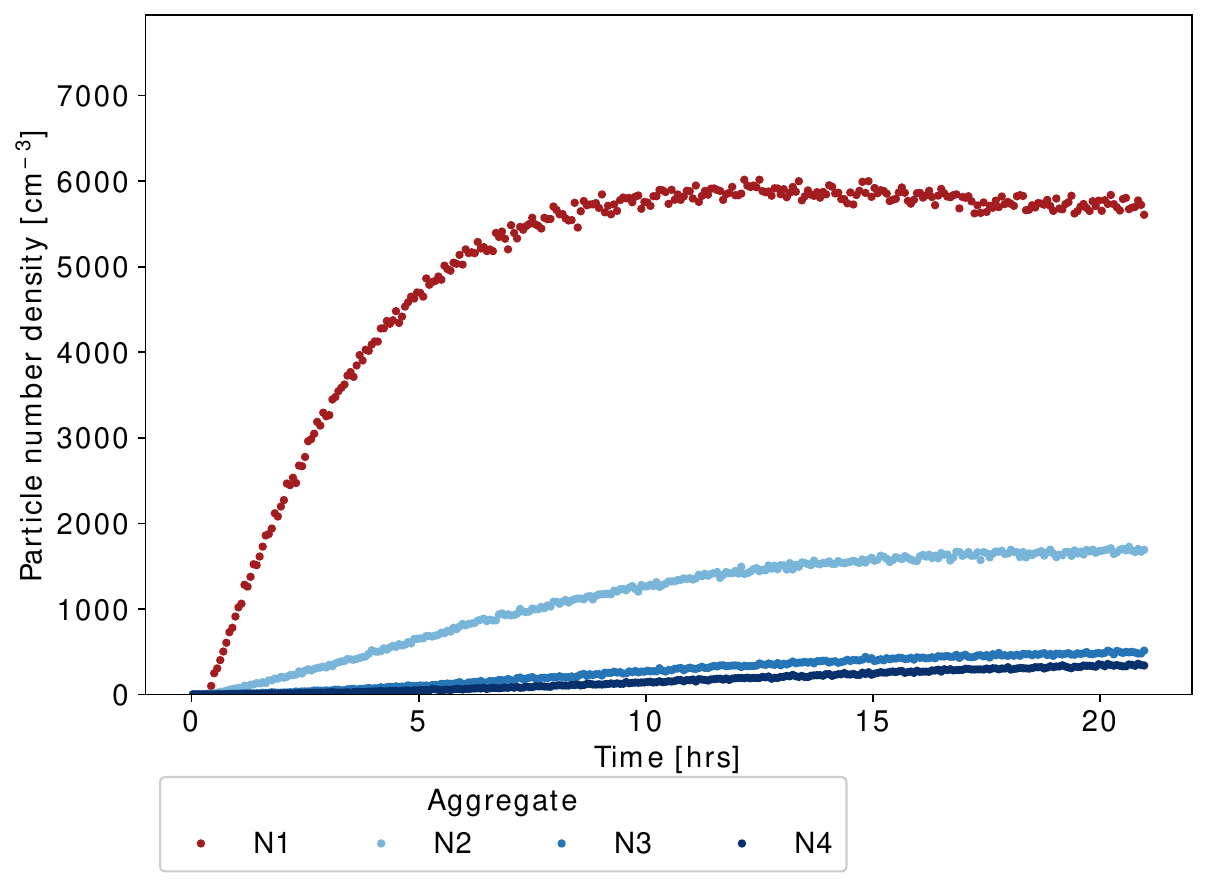}
     \caption{Number densities of single particles, N1, and each of the aggregates, N2-N4, over time. Shows experiment nr. 9 in Tab. \ref{tab:setup}}
    \label{fig:aggregation}
\end{figure}

\begin{figure}
    \centering
    \includegraphics[width=\linewidth]{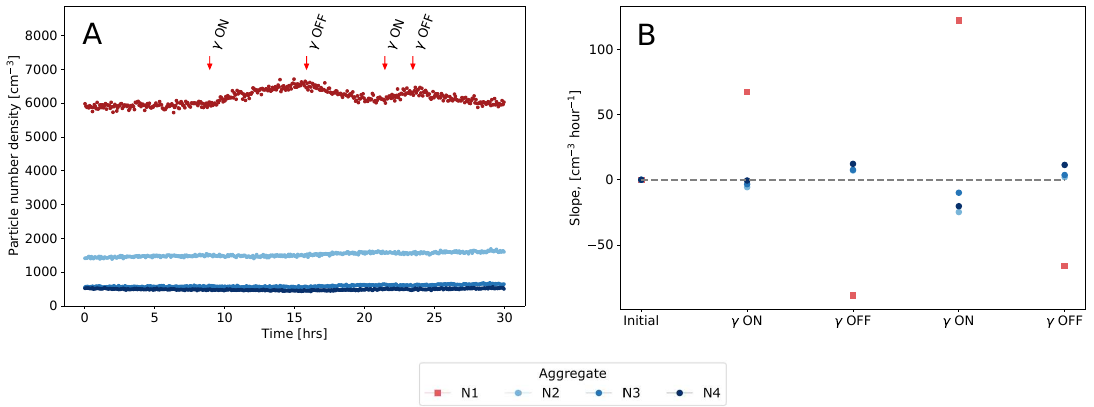}
    \caption{Effect of gamma radiation on the aggregation of particles in a low-humidity environment (RH $\approx$ 20\%). A) Number density of single particles, N1, and each of the aggregates, N2-N4, over time . Arrows indicate the presence of absence of gamma radiation. B) The change in number density found as the slope of linear fits for each radiation setting shown in (A) in relation to the initial slope. Each linear fit is started 30 min after the change in radiation (gamma being turned on or off), in order to allow for ionization to take effect, and ended just before the next change. Dashed line at 0.0 indicates no change in particle number. Show experiment nr. 10 in Tab. \ref{tab:setup}}
    \label{fig:aggregation_both}
\end{figure}

As shown in Fig. \ref{fig:size_both}, the original d $\approx$ 50 nm particles (N1) introduced into the chamber aggregate to form larger clusters of two to four particles (N2-N4). In order to assess the aggregation of the particles, Fig. \ref{fig:aggregation} shows the number of each aggregate over time, as defined from the full width half maximum of the peaks in the size distributions. As can be seen from the figure the number density of the N1 particles increases steadily up to $\sim$10 hours after the introduction of the particles. In comparison the number of larger particles keeps increasing even after the curve for N1 particles have reached a plateau. 

In order to study the effect the gamma radiation on the aggregation of the particles we looked at the change in the production rate of the particles before and after gamma radiation was introduced. Fig. \ref{fig:aggregation_both}A, shows how the number of each aggregate changed with the introduction of gamma radiation. To get a clearer view of especially the larger aggregates linear curves were fitted to the data points shown in Fig. \ref{fig:aggregation_both}A, after each change in gamma radiation. The change in the slopes of each linear fit can be seen in Fig. \ref{fig:aggregation_both}B. Linear fits were made for each change in the gamma radiation, corresponding to the sections between the arrows in Fig. \ref{fig:aggregation_both}A and the sections before and after the first and last arrow respectively. Each linear fit was started 30 min after the change in radiation in order to allow for the ionization of the gas in the chamber to adapt to the change. The values plotted in Fig. \ref{fig:aggregation_both}B are calculated as the slopes of these linear fits subtracted from the initial slope before the first change in radiation. Here we notice a trend for gamma radiation to cause an increase in the number of single particles (N1) and a decrease in the number of aggregates (N2-N4). Similarly, turning the gamma radiation off caused a decrease in the number of single particles (N1) and an increase in the number of aggregates (N2-N4). This indicates that gamma radiation inhibits the aggregation of the particles.

\subsection{Charging state} \label{sec:charge}  

\begin{figure*}
    \resizebox{\hsize}{!}
    {\includegraphics[width=0.8\linewidth]{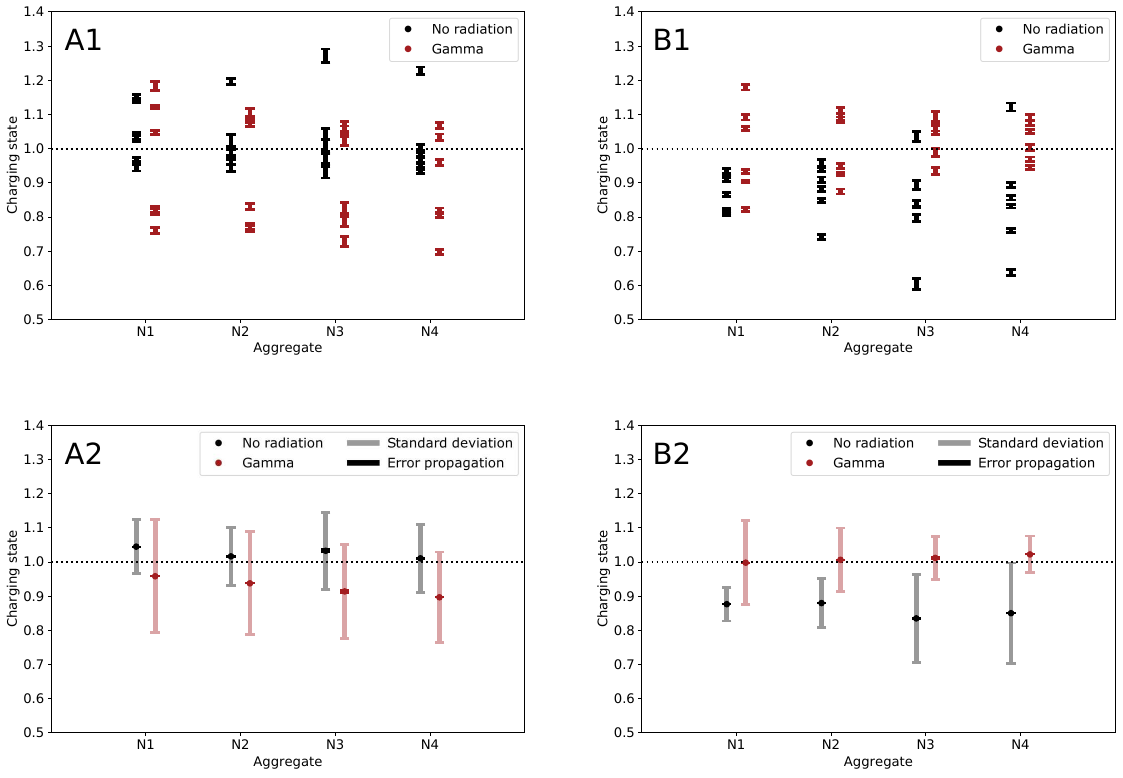}}
    \caption{Charging state of positively (A) and negatively (B) charged particles. A1 and B1 show charging state of all repeated experiments with error bars indicating propagated errors. A2 and B2 show average charging states for each aggregate, with dark error bars indicating propagated errors (notice that they are so small that they fit within the data points), and light error bars indicating standard deviation among repeated experiments. Horizontal dotted line at 1.0 indicates bipolar charge distribution. Show experiments nr. 1-8 in Tab. \ref{tab:setup}}
    \label{fig:charging_state_all}
\end{figure*}

As explained in Sec. \ref{sec:SMPS} the charging state is a measure of the fraction of charged particles in the population compared to the fraction of charged particles in a steady-state bipolar distribution. As such the charging state indicates the charging relative to a standard distribution, and potential changes in the charging state due to irradiation with gamma radiation can indicate how high-energy radiation affects the overall charging of the particle population. 

Fig. \ref{fig:charging_state_all} shows the charging state, individually and averaged, of all experiments of positively and negatively charged particles respectively, Each plot shows the charging state for both non-irradiated particles, and particles irradiated with gamma radiation. The experiments plotted are listed in Tab. \ref{tab:setup}. Experiments with and without bipolar charger are done consecutively, and charging states are calculated for each pair. A total of six individual charging state measurements are plotted in A1 and A2 and used to find each average shown in the figures A2 and B2.

For the non-irradiated case, the positively charged particles in Fig. \ref{fig:charging_state_all}A2 have charging states just above the steady-state distribution, as illustrated by the horizontal dotted line. This corresponds to a slight overcharging of the particles, and indicates that there are more positively charged particles than expected from the steady-state distribution. Gamma radiation decreases the number of positively charged particles, thereby lowering the charging state, resulting in the particle population becoming undercharged compared to the steady-state distribution.

For the non-irradiated case, the negatively charged particles in Fig. \ref{fig:charging_state_all}B2 generally have charging states below the horizontal line, indicating a lack of negatively charged particles in the population compared to the steady-state distribution. Gamma radiation increases the number of negatively charged particles, bringing the charging state close to the steady-state distribution. 

Both plots' averages in Fig. \ref{fig:charging_state_all} indicate a shift towards more negatively charged particles in the population as a result of the gamma radiation. 

\subsubsection{Statistical T-test of charging states}

In order to test the effect of gamma radiation on the particles we compare the mean charging state of irradiated and non-irradiated particles. Since the test needed would be a comparison of the means of two independent samples, an independent T-test was chosen. 
The test was performed for positively and negatively charged particles separately, and the samples tested correspond to the data plotted in Fig. \ref{fig:charging_state_all} A1 and B1. 

\begin{table}
    \centering
    \begin{tabular}{c|c|c|c|c}
        & N1 & N2 & N3 & N4 \\
        Positively charged & 0.31 & 0.33 & 0.16 & 0.16 \\
        Negatively charged  & 0.07 & 0.04 & 0.02 & 0.03 
    \end{tabular}
    \caption{P-values for t-test of two independent samples comparing irradiated and non-irradiated particles. Low-humidity environment, Fig. \ref{fig:charging_state_all}.}
    \label{tab:t-test}
\end{table}

The probability values (p-values) from the T-test are shown in Tab. \ref{tab:t-test}. For the positively charged particles we find p-values that are all outside the a significance level of 5\% (i.e. all p-values are above 0.05). For the negatively charged particles, most of the aggregates have p-values within a significance level of 5\%, the only exception from this is the N1 particles showing a p-value of 7\%.
These results indicate that the effect of gamma radiation seen for the average charging states shown in Fig. \ref{fig:charging_state_all} is statistically significant for the negatively charged particles (B2), but not for the positively charged particles (A2).

\subsection{Relative humidity}\label{sec:highhum}

\begin{figure*}[h!]
    \resizebox{\hsize}{!}
    {\includegraphics[width=0.8\linewidth]{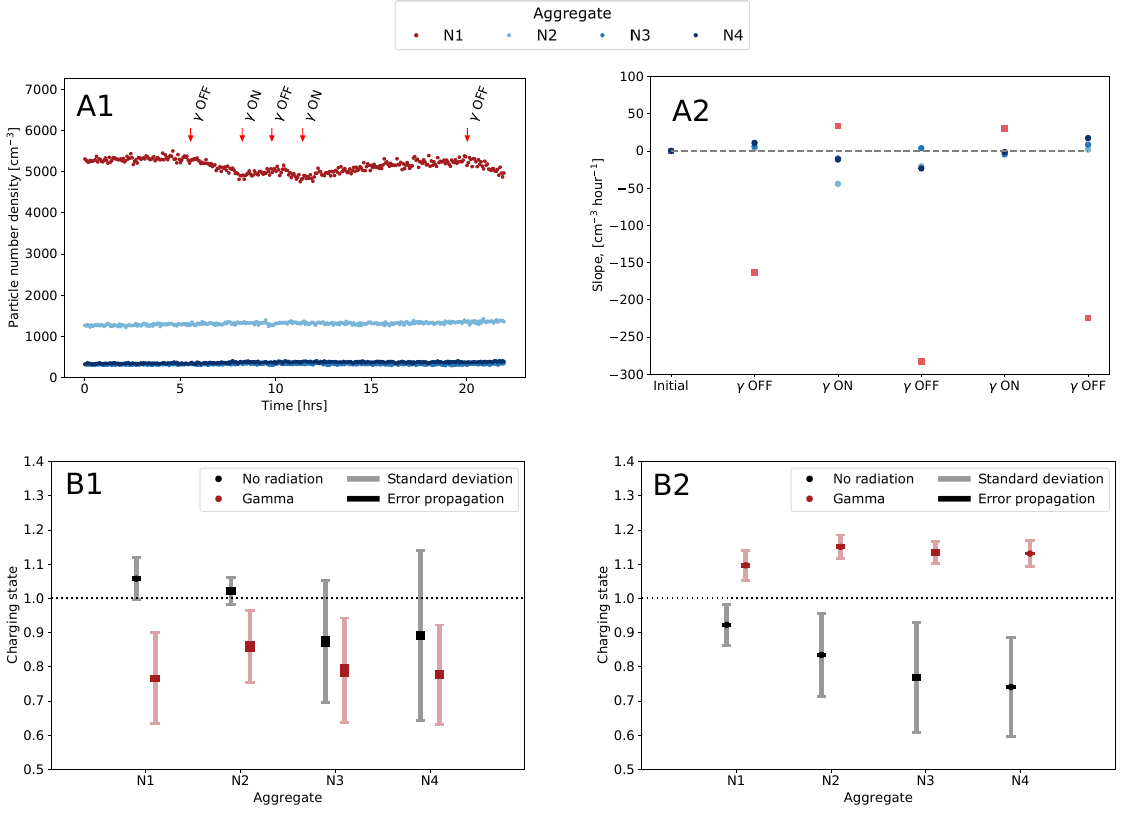}}
    \caption{Experiments conducted in a high humidity environment (RH $>$ 50\%).
    A) Number densities of aggregates over time. The change in density caused by gamma radiation is shown in A2 as slopes from linear fits for the different radiation settings as marked by arrows in A1. Experiment nr. 19 in Tab. \ref{tab:setup}. B) Average of the charging states calculated for each of the repeated experiments (nr. 11-18 in Tab. \ref{tab:setup}). B1 shows the positively charged particles, and B2 shows the negatively charged particles. Dark error bars indicate propagated errors (notice that they are so small that they often fit within the data points), and light error bars indicate standard deviation among repeated experiments.}
    \label{fig:high_all}
\end{figure*}

As mentioned in Sec. \ref{sec:methods} the lowest relative humidity reached in the chamber was RH $\approx$ 20\%, and all measurements in the previous sections were done at this RH value.
In order to test the effect of relative humidity on the aggregation and charging of the particles a set of experiments were repeated at higher relative humidities with RH $>$ 50\%. The experimental settings can be seen in Tab. \ref{tab:setup} (nr. 11-19), and the results are shown in Fig. \ref{fig:high_all}.
The top row of the figure, A1 and A2, shows that gamma radiation also inhibits the aggregation of the particles in a high-humidity environment, similarly to the low humidity cases in Fig. \ref{fig:aggregation_both}. Fig. \ref{fig:high_all}A, show a trend of an increase in the number of N1 particles under irradiation with gamma, similarly to what was seen for the low-humidity case. The changes observed for the N1 particles shown in Fig. \ref{fig:high_all}A2 are even more pronounced than those seen for the low-humidity case in Fig. \ref{fig:aggregation_both}B, with average changes in slopes of $\sim$250 $\pm$ 60 cm$^{-3}$hr$^{-1}$ in high humidity compared to $\sim$170 $\pm$ 20 cm$^{-3}$hr$^{-1}$ in low humidity.
The trend for the aggregates (N2-N4) is less clear in this high-humidity case compared to the low-humidity. 
The bottom row of Fig. \ref{fig:high_all}, B1 and B2, show the average charging states of positively and negatively charged particles respectively, corresponding to the low-humidity cases, Fig. \ref{fig:charging_state_all}A2 and Fig. \ref{fig:charging_state_all}B2.
The positively charged particles in Fig. \ref{fig:high_all}B1 show a general trend for the irradiated particles to be more undercharged than the non-irradiated particles. This indicates that the number of positively charged particles decreases under gamma radiation. The opposite trend is seen for the negatively charged particles in Fig. \ref{fig:high_all}B2, where the irradiated particles are significantly more overcharged compared to the non-irradiated particles indicating an increase in the number of negatively charged particles under gamma radiation. Both these trends are similar to what was seen in the low-humidity environment, where gamma radiation was observed to shift the particles population towards a more negative charge.

To test the statistical significance of the effect of gamma radiation on the charging state in a high-humidity environment a T-test was performed, and the resulting p-values can be seen in Tab. \ref{tab:t-test_high}. For the positively charged particles it is found that the effect of gamma radiation is not statistically significant for most of the aggregates (the exception being the N1 particles). For the negatively charged particles it is found that the effect of gamma radiation is statistically significant for all aggregates, with p-values that are well within the 5\% significance level. 
The fact that the changes caused by gamma radiation are more statistically significant in the high-humidity environment, could indicate that the increase in the number of negatively charged particles is more pronounced at higher humidity. 

\begin{table}
    \centering
    \begin{tabular}{c|c|c|c|c}
        & N1 & N2 & N3 & N4 \\
        Positively charged & 0.05 & 0.11 & 0.64 & 0.60 \\
        Negatively charged  & 0.03 & 0.02 & 0.04 & 0.02
    \end{tabular}
    \caption{P-values for t-test of two independent samples comparing irradiated and non-irradiated particles. High-humidity environment, Fig. \ref{fig:high_all} B1 and B2.}
    \label{tab:t-test_high}
\end{table}

\section{Discussion} \label{sec:discussion}

If we wish to understand exoplanets in different planetary systems it is important to understand how the radiation field of the host star influences the exoplanet atmospheres. Here high-energy radiation plays an important role. 

The radiation environment caused by the host star and galactic environment has been found to have an impact on both dynamics, chemistry, and climate of exoplanets (reviewed by e.g. \cite{airapetian_impact_2020}). High-energetic particles, such and cosmic rays and SEPs, have been found to affect both the chemical composition \citep{barth_moves_2021} and ionization \citep{rodriguez-barrera_environmental_2018} of the atmospheres. Due to the importance of ionization and charging in ion-induced nucleation \citep{wagner_role_2017}, high-energetic particles might play a significant role in cloud formation in environments with low presence of primary aerosols \citep{jokinen_ion-induced_2018}, such as gas planets.

The majority of all stars in the Milkyway are M-dwarfs, and given our current observation biases, these make for interesting targets for transit observations of exoplanets in the habitable zone \citep{mesquita_earth-like_2021}. Due to the low luminosity of M-dwarfs their habitable zone is very closer to the star \citep{kasting_habitable_1993,selsis_habitable_2007}, which allows for smaller planets to be observable in the habitable zone. 
M-dwarfs have been found to remain magnetically active for longer than Sun-like stars \citep{west_spectroscopic_2004,scalo_m_2007,guinan_living_2016}, which creates a very different radiative environment than in the Solar System. The high activity causes the stars to have stronger flares \citep{vida_frequent_2017,tilley_modeling_2019}, and release more stellar energetic particles and gamma radiation \citep{griemeier_cosmic_2005,obridko_young_2020,sadovski_cosmic_2018,fraschetti_stellar_2019}. 
Exoplanets in these systems might therefore be exposed to a higher degree of high-energy radiation from the host star, both due to the higher activity and due to the potentially closer proximity to the star.

The exact correlation between the activity of the host star and the total influx of high energy radiation to an exoplanet is, however, nontrivial. While the particle influx from the star itself increases with stellar activity, the corresponding increase in magnetic flux and stellar wind has also been found to shield the system from the influx of GCR from outside the system (\cite{mesquita_earth-like_2021,mesquita_galactic_2021,nandy_solar_2021}, and reviewed in e.g. \cite{potgieter_solar_2013}). The nature of the high-energy particles in a system will therefore change depending on the activity of the host star during its evolutionary phases, such that SEPs might be dominating during the younger active phase of the star, while GCRs will be dominating later in the evolution or for less active stars \citep{rodgers-lee_stellar_2021}. \\

In this study we imitate the high-energy radiation of SEPs and GCRs with gamma radiation, similarly to what has been done previously \citep{enghoff_measurement_2017}. Gamma radiation is one of the products released in the interaction between SEPs and GCRs and celestial atmospheres, and high-energy particles will therefore release gamma radiation both from the host star and from the atmosphere of the planet itself, making the Earth atmosphere a bright source of gamma radiation (as reviewed by e.g. \cite{mazziotta_cosmic-ray_2020} and \cite{dean_gamma-ray_1988}). The Cs-137 sources used in this study release gamma radiation with energies of 0.66 MeV which lies within the energy range of SEPs and low-energy GCRs (e.g. \cite{barth_moves_2021}) as well as the range of cosmic ray showers caused by higher energy GCRs. The duration of each experiment has been determined by the scan times and number of desired consecutive measurements, and the chamber has therefore been irradiated with gamma radiation for periods of $\sim$4 hours at a time. This is longer than the average SEP event, but still lies well within the general time scales from minutes to hours (e.g. \cite{firoz_duration_2022,doyle_investigating_2018,reep_what_2019}),while the influx of GCRs is more continuous. It should be noted that studies have found that in the case of ion-induced nucleation the nature of the ionizing radiation is not important as long as the ionizing effect is the same \citep{enghoff_aerosol_2011}. \\

The aim of this paper has been to study how the aggregation and charging of mineral cloud particles are affected by high-energy radiation and humidity. Using experiments with SiO$_2$ particles in an atmosphere chamber we observed two trends: 1) gamma radiation inhibits the aggregation of the particles, and 2) gamma radiation causes a higher number of particles to become negatively charged. 
These trends, of less aggregation and more negative charge, are observed both in environments with lower relative humidity (RH $\approx$ 20\%) and higher relative humidity (RH $>$ 50\%). \\

The coupling between the effects of gamma radiation (i.e. the inhibited aggregation and the negative charge of the particles) is uncertain, but it might be related to a change in the mobility of the particles. As can be seen from Fig. \ref{fig:size_pn}, the mobility size of the negatively charged particles is slightly larger than for the positively charged particles. As the mobility diameter is linked to the size of the particles, as described in Sec. \ref{sec:methods}, this indicates a slight difference in the size of the positively and negatively charged particles with the negatively charged particles being larger. Since the gamma radiation leads to an increase in the number of negatively charged particles it leads to an overall increase in particle size in the chamber. At these particle sizes the coagulation coefficient between similar sized particles decreases with increasing particles size \citep{fujimoto_langevin_2021,fuchs1964mechanics} leading to fewer collisions overall in the chamber. Thus the decrease in aggregation might be linked to a decrease in the number of collisions caused by a lower mobility of the negatively charged particles. It is uncertain what causes the difference in mobility diameter between negatively and positively charged particles in this study. \\

The effect of gamma radiation on the charging state, as shown in Fig. \ref{fig:charging_state_all}B and \ref{fig:high_all}B, indicates that gamma radiation causes a decrease in the charging state for positively charged particles and an increase for negatively charged particles for both the low-humidity and high-humidity case. However, this change is only found to be statistically significant for the negatively charged particles and not the positively charged particles. It is yet unknown whether the decrease in the number of positively charged particles would get more statistically significant with more repetitions of the experiments, or it is caused by an actual physical process allowing the particles to maintain their positive charge under gamma radiation. 

The overall negative charging of the particles might be linked to the mobility of the ions created by the gamma radiation. In Earth's atmosphere, high-energy radiation, such as SEPs and GCRs, lead to the formation of ions and ionic clusters through the separation of electrons from N$_2$ or O$_2$, that will subsequently be captured by neutral molecules (reviewed in \cite{harrison_ion-aerosol-cloud_2003}). Depending on the composition of these ions, they will have different mobility and thereby different rate of uptake by aerosols. In Earth's troposphere, the negative ions generally have higher mobilities than the positive (e.g. \cite{horrak_bursts_1998,harrison_ion-aerosol-cloud_2003}), and some of the dominating negative ions are O$_2 ^-$(H$_2$O)$_n$ and NO$_3^-$(H$_2$O)$_n$, while some of the dominating positive ions are H$_3$O$^+$(H$_2$O)$_n$, H$^+$(H$_2$O)$_n$, and NO$^+$(H$_2$O)$_n$. Most of these ions could also be present in our experiments, since they can be formed through ionisation of N$_2$ and H$_2$O (the gas composition used in this study), and the Cs-137 gamma sources we use can ionize both N$_2$ and H$_2$O. This could indicate that we might also expect a higher mobility of the negative compared to the positive ions in our setup, similar to what is seen in Earth's atmosphere. A noticeable difference between the gas composition in this study and Earth's atmospheric composition is the lack of O$_2$ in the chamber. A previous study by \citep{wiedensohler_investigation_1986}, looked at the charging distribution of positively and negatively charged particles in different gas composition. They found a larger split in the charging state of positively and negatively charged particles in N$_2$ gas compared to air, with a significantly higher number of negatively charged particles. They suggest that this might be due to the lack of O$_2$ since this results in a lack of electron receptors, leading to more free electrons and thereby potentially a higher average mobility of all negative ions. Exactly how the number and mobility of the small air ions will affect the final charging of particles is non-trivial, and depends on a series of chemical and ion exchange reactions where ion clusters build up before they collide with and are absorbed by the larger particles \citep{harrison_ion-aerosol-cloud_2003}. But as observed by \cite{wiedensohler_investigation_1986} the initial composition of air ions can affect the final charging. \\

Different degrees of relative humidity were tested in this study, and as mentioned, gamma radiation is shown to have a similar effect on the SiO$_2$ particles in low- and high-humidity environments. Previous studies have suggested that water molecules structure themselves differently on silicates, depending on the relative humidity of the environment \citep{asay_evolution_2005}. Based on the terminology from \cite{asay_evolution_2005}, the two humidity ranges we are observing lead to structures that are ice-like (RH = $\sim$ 20\%) and transitional/liquid (RH $>$ 50\%) respectively. The fact that we observe similar effects of gamma radiation in the two cases, could indicate that the charging and inhibition of aggregation is not dependent on the structuring of water molecules on the surface of the particles. However, the adhesion of water molecules to solid surfaces (reviewed in e.g. \cite{zhou_review_2021,sacchi_water_2023}), and the resulting effect on charging and aggregation of particles \citep{he_atmospheric_2019,he_charging_2020}, is complex and depends on both the physical and chemical properties of the surfaces, and we have not accounted for it in this study. \\

Based on the results of this and previous studies we suggest that high-energy environments might promote the formation of smaller cloud particles.
Previous studies have shown that high-energy radiation induces the nucleation of cloud particles in Earth-like atmospheres, thereby increasing the number of new particles \citep{lee_new_2019,wagner_role_2017,svensmark_response_2013}. In this study, we show that high-energy radiation may inhibit aggregation of the existing particles, thereby favouring a high number of smaller particles rather than a low number of larger particles. Together this could indicate that high-energy radiation increases the number of particles overall. Under the right conditions, these particles might act directly as cloud particles or potentially as cloud condensation nuclei for other cloud forming species. For Earth, it has been speculated that the required size of a cloud condensation nuclei is 50 nm (e.g. 
\cite{svensmark_response_2013,sarangi_simplified_2015,fan_substantial_2018}), and under Earth-like conditions, previous studies \citep{svensmark_response_2013} have found that gamma radiation promotes the growth of nucleated particles up to $>$ 50 nm. Since this study has found that gamma radiation inhibits aggregation of d $\approx$ 50 nm particles, this could indicate that high-energy radiation will increase the number of cloud condensation nuclei overall by promoting their formation and preventing the loss in number through aggregation. We note that the studies promoting nucleation and growth of smaller aerosols were performed on a different set of species than in this study.\\

With telescopes such as James Webb Space Telescope (JWST) our understanding of cloud formation is being put to the test as it becomes essential in our analysis of transit spectra. Already the early release observations from JWST \citep{pontoppidan_jwst_2022} presented planets such as WASP-96b with strong H$_2$O features, that are also predicted to have mineral cloud particles including silicate species ($\geq$40\%) \citep{samra_clouds_2023}, indicating the relevance of understanding the behaviour of mineral cloud particles in humid environments. The presence of mineral cloud particles is not unusual, and has been observed for e.g. WASP-107b both with Hubble Space Telescope \citep{kreidberg_water_2018} and later with JWST (preprint \cite{dyrek_so2_2023}). Another potential JWST target of interest is HD189733b, that is both expected to have silicate cloud particles \citep{lee_dynamic_2016,barth_moves_2021}, and where observations have shown flares from the host star \citep{bourrier_moves_2020}, indicating that an increased level of gamma radiation might be present in the upper atmosphere of the planet. These observations and models all show the importance of understanding the interplay between mineral cloud particles and high-energy radiation. \\  

The results of this study have been primarily qualitative and solely focused on SiO$_2$ as an analogue to mineral cloud particles. In order to get a better understanding of the behaviour of mineral cloud particles in general, it would be highly beneficial to conduct further extensive studies with more repetitions and other predicted cloud formation species, such as silicate and magnesium species, like enstatite (MgSiO$_3$) and forsterite (Mg$_2$SiO$_4$) (e.g. \cite{lunine_evolution_1986,helling_dust_2008,carone_wasp-39b_2023,gao_universal_2021}), or titanium species such as TiO$_2$ (e.g. \cite{kohn_dust_2021,helling_comparison_2008}).


\section{Summary} \label{sec:Con}
In this study we have explored the question of how high-energy radiation affects the aggregation and charging of mineral cloud particles.
We here present the results of a series of experiments with mineral SiO$_2$ particles in an atmosphere chamber under varying degrees of high-energy radiation and relative humidity. 

Based on the experiments we observe the following:
\begin{enumerate}
    \item The d $\approx$ 50 nm SiO$_2$ particles (N1) aggregate in the chamber to form clusters of two, three, and potentially four particles (N2 - N4).
    \item Gamma radiation inhibits the overall aggregation of the SiO$_2$ particles. This is observed as an increase in the number of single particles (N1) and a decrease the number of larger aggregates (N2 - N4). It is unknown if the trend reflects a prevention of the N1 particles to aggregate, or if it reflects a fragmentation of existing aggregates (N2-N4), but it could be linked to a lowered mobility of the N1 particles due to a change in their charging state.
    \item Gamma radiation causes the particle population to become more negatively charged. This is observed as an increase in the charging state of the negatively charged particles and a decrease in the charging state of positively charged particles, when gamma radiation is present. This change in charging state is only found to be statistically significant for negatively charged particles.
    \item The effect of gamma radiation is visible both at lower ($\sim$20\%) and higher ($\geq$50\%) relative humidity.
\end{enumerate}

We suggest that gamma radiation may favour the formation of many smaller particles based on: 1) previous studies have found that gamma radiation increases ion-induced nucleation, and 2) this study has found that gamma radiation decreases aggregation. 
Depending on the environment, the many smaller particles might act directly as cloud particles or as cloud condensation nuclei for other cloud forming species.

\begin{acknowledgements}
The authors thank Matthew S. Johnson for insightful discussions on the background for the study and for the loan of instruments.
This project is  funded by the European Union’s Horizon 2020 research and innovation programme under the Marie Sklodowska-Curie grant No 860470. This project received funding from St Leonard’s Postgraduate College from the University of St Andrews.
\end{acknowledgements}

\bibliography{references}
\bibliographystyle{aasjournal}

\begin{appendix}

\section{Loss rate to chamber} \label{app:loss}
\begin{figure}[H]
    \centering
    \includegraphics[width=1\linewidth]{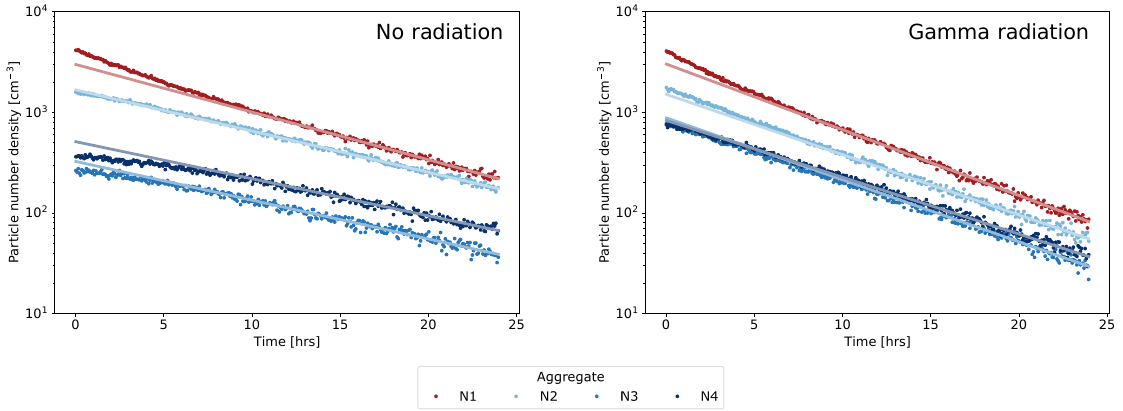}
    \caption{Number density of single particles, N1, and each of the aggregates, N2-N4, over time after cutting off the flow of particles into the chamber. Left: The chamber is not irradiated. Right: The chamber is irradiated with gamma radiation. The lines show exponential curves fitted to the data points for the last 10 hours for each of the aggregates.}
    \label{fig:loss}
\end{figure}

Over time, particles will be lost to the chamber walls due to collisions.
In order to investigate how many particles are lost to the walls, we study the decrease in particle number density after ending the input of additional particles. The results can be seen in Fig. \ref{fig:loss}.

The figures show the total decrease in particle number density over a time period of 25 hours after the atomizer has been turned off, for non-irradiated and irradiated particles respectively. 
The number density of each of the aggregates can decrease due to three processes: 1) Aggregation leads to a decrease of one aggregate population, and an increase of an other population. 2) Particles are lost with the airflow, e.g. to the measurements. 3) Particles collide with the chamber walls and stay there. 

If the particles were solely lost to the airflow and the chamber walls, we would expect the decrease to be exponential for each of the aggregates. 
Exponential curves have been fitted to each of the aggregates for the last 10 hours (hour 15 to 25). 
The plots in Fig. \ref{fig:loss} both show that the number density of smaller aggregates (N1 and partly N2) initially decreases faster than exponential, whereas the density of the larger aggregates initially decreases slower than exponential, which confirms that aggregation continues to take place after the input of new particles has stopped. For the last 10 hours, where the exponential curves have been fitted, the aggregates follow these fits with R$^2$-values of $\geq$0.9. 
Based on the fact that the number densities follow an exponential fit for the last 10 hours, we assume that aggregation is insignificant for this time span. Following this assumption the exponent for the exponential fits should reflect the loss of particles solely due to the airflow and the chamber walls. 
Notice that the input airflow was different for the two cases (13 L/min for the case without radiation and 16 L/min for the case with gamma radiation), leading to the particles being diluted faster for the irradiated case. 

Given that the exponential loss of particles follow the curve:
\begin{equation}
    N(t) = N_0 * exp(-a \cdot t)
\end{equation}
Where N(t) is the particle number density at time, t, and a is the loss rate of the particles. The total loss rate of the particles can be expressed as:
\begin{equation}
    a = a_{airflow}+a_{walls} = flow_{air}\cdot V_{chamber}^{-1} + a_{walls}
\end{equation}

The loss rate of particles lost to the airflow (a$_{airflow}$) is found as the total flow of air through the chamber (flow$_{air}$) over the chamber volume (V$_{chamber}$). Values can be seen in Tab. \ref{tab:loss}. 

In order to find the total particle loss rate to the chamber walls, the particle number densities shown in Fig. \ref{fig:loss} have been added up to find the total number density in the chamber (N1+N2+N3+N4). Exponential curves were fitted to the last 10 hours similarly to what is shown in Fig. \ref{fig:loss}, and the loss rates to the chamber walls was found from the exponent and the loss rate to the airflow.

\begin{table}[H]
    \centering
    \begin{tabular}{c|c|c|c}
        & Total loss rate & Loss rate to airflow & Loss rate to chamber walls \\
        No radiation & -0.100 & -0.098 & -0.002 \\
        Gamma radiation & -0.142 & -0.120 & -0.022
    \end{tabular}
    \caption{Loss rates     of total number of aggregates in units of hr$^{-1}$.}
    \label{tab:loss}
\end{table}

\section{Full mobility diameter range}

\begin{figure}[H]
    \centering
    \includegraphics[width=0.6\linewidth]{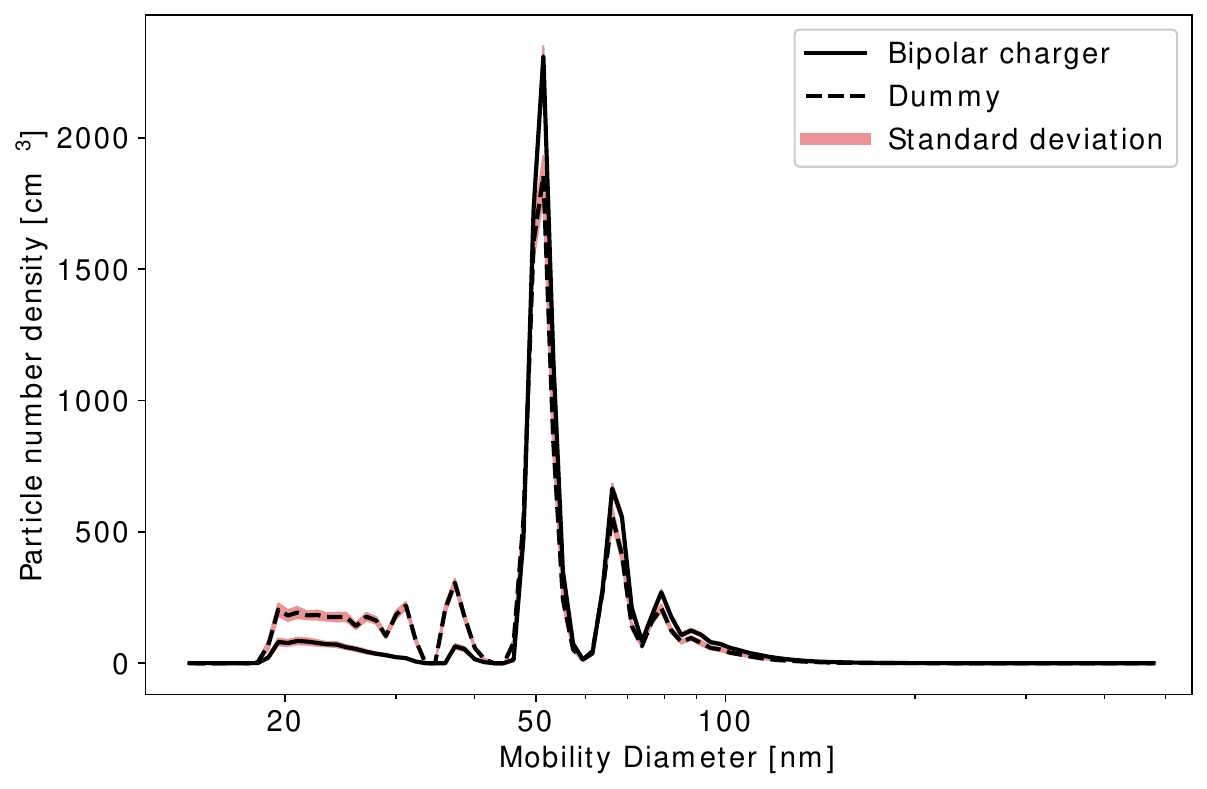}
    \caption{Size distribution of negatively charged particles measured with and without bipolar charger. Red shading indicates standard deviation among the 50 measurements constituting each experiment. Figure shows full diameter range of Fig. \ref{fig:size_both} and displays the results of one of experiment nr. 5 in Tab. \ref{tab:setup}. Features at lower mobility diameters are believed to be caused by aerosols released by the the Milli-Q water (e.g. \cite{belosi_comparison_2013,marcolli_ice_2016}).}
    \label{fig:size_full}
\end{figure}

\end{appendix}

\end{document}